\documentclass
[aps,pra,twocolumn,nofootinbib,a4paper,floatfix,superscriptaddress,10pt]{revtex4-1}%
\usepackage{graphics}
\usepackage{amsmath}
\usepackage{graphicx}
\usepackage{amsfonts}
\usepackage{amssymb}
\usepackage{float}
\usepackage{longtable}
\usepackage{epsfig}
\usepackage{latexsym}
\usepackage{theorem}
\usepackage{bbm}
\usepackage{bm}
\usepackage{verbatim}
\usepackage{psfrag}
\usepackage{color}
\usepackage{dsfont}
\usepackage{subfigure}
\usepackage{verbatim}
\usepackage{dcolumn}
\usepackage{bm}
\usepackage[dvipsnames]{xcolor}
\usepackage{listings}
\usepackage{mathrsfs}
\usepackage{filecontents}
\usepackage{soul}
\usepackage{times}

\newcommand*\mean[1]{\bar{#1}}
\makeatletter
\def\maketag@@@#1{\hbox{\m@th\normalfont\normalsize#1}}
\makeatother
\begin{document}

\title{Tight bounds for private communication over bosonic Gaussian channels\\based on teleportation simulation with optimal finite resources}

\author{Riccardo Laurenza}
\affiliation{Department of Computer Science, University of York, York YO10 5GH, United Kingdom}
\affiliation{QSTAR, INO-CNR and LENS, Largo Enrico Fermi 2, 50125 Firenze, Italy}
\author{Spyros Tserkis}
\affiliation{Centre for Quantum Computation and Communication Technology, School of Mathematics and Physics, University of Queensland, St Lucia, Queensland 4072, Australia}
\author{Leonardo Banchi}
\affiliation{Department of Physics and Astronomy, University of Florence, via G. Sansone 1, I-50019 Sesto Fiorentino (FI), Italy}
\author{Samuel L. Braunstein}
\affiliation{Department of Computer Science, University of York, York YO10 5GH, United Kingdom}
\author{Timothy C. Ralph}
\affiliation{Centre for Quantum Computation and Communication Technology, School of Mathematics and Physics, University of Queensland, St Lucia, Queensland 4072, Australia}
\author{Stefano Pirandola}
\affiliation{Department of Computer Science, University of York, York YO10 5GH, United Kingdom}
\affiliation{Research Laboratory of Electronics, Massachusetts Institute of Technology (MIT), Cambridge, Massachusetts 02139, USA}

\begin{abstract}
Upper bounds for private communication over quantum channels can be derived by
adopting channel simulation, protocol stretching, and relative entropy of
entanglement. All these ingredients have led to single-letter upper bounds to
the secret key capacity which can be directly computed over suitable resource
states. For bosonic Gaussian channels, the tightest upper bounds have been
derived by employing teleportation simulation over asymptotic resource states,
namely the asymptotic Choi matrices of these channels. In this work, we adopt
a different approach. We show that teleporting over an analytical class of
finite-energy resource states allows us to closely approximate the ultimate
bounds\ for {increasing energy}, so as to provide increasingly tight upper
bounds to the secret-key capacity of one-mode phase-insensitive Gaussian
channels. We then show that an optimization over the same class of resource
states can be used to bound the maximum secret key rates that are achievable
in a finite number of channel uses.

\end{abstract}
\maketitle


\section{Introduction}

The ultimate performance of a communication channel is given by its capacity.
In quantum information theory~\cite{Watrous,Hayashi,NielsenChuang,Bengtsson},
there are several definitions of capacity, depending on whether one wants to
send classical information, quantum information, entanglement etc. In
particular, the secret-key capacity of a quantum channel represents the
maximum number of secret bits that two authenticated remote users may extract
at the ends of the channel, without any restrictions on their local operations
(LOs) and classical communication (CC), briefly called LOCCs. This capacity is
particularly important because it upper-bounds the secret key rate of any
point-to-point protocol of quantum key distribution (QKD)~\cite{BB84,Ekert}
(see Ref.~\cite{QKDreview} for a comprehensive review). In this context, the
highest key rates are those achievable by QKD\ protocols implemented with
continuous-variable (CV) systems, i.e., bosonic modes of the electromagnetic
field, which are conveniently prepared in Gaussian
states~\cite{Weedbrook.et.al.RVP.12,Adesso.Ragy.OSID.14,SamRMPm,Serafini.B.17,hybrid}%
. These quantum states are transmitted through optical fibers or free-space
links which are typically modeled as one-mode Gaussian
channels~\cite{Weedbrook.et.al.RVP.12,Holevo.PIT.07,Caruso,HolevoVittorio} ,
to be considered as the direct effect of collective Gaussian
attacks~\cite{GaussianAttacks}.

Exploring the ultimate achievable rates of
CV-QKD~\cite{CVQKD1,CVQKD2,CVQKD3,CVQKD4,CVQKD5,CVQKD6,CVQKD7,CVQKD8,CVMDIQKD}
has been a very active research area. Back in 2009, a lower bound to the
secret key capacity of the thermal-loss channel was given~\cite{ReverseCAP} in
terms of the reverse coherent information~\cite{RevCohINFO,Ruskai}. For a
pure-loss channel of transmissivity $\tau$, this work established that the
rate of an optimal point-to-point QKD protocol can achieve a linear scaling of
$1.44~\tau$ bits per channel use. In 2014, a (non-tight) upper bound was found
by resorting to the squashed entanglement~\cite{TGW}, confirming the $\sim
\tau$ scaling in a pure-loss channel. More recently, a tighter and definitive
upper bound has been established by Ref.~\cite{Pirandola.et.al.NC.17} in terms
of the relative entropy of entanglement
(REE)~\cite{Vedral.RMP.02,Vedral.et.al.PRL.97}. For a pure-loss channel, the
lower and upper bounds of Refs.~\cite{ReverseCAP,Pirandola.et.al.NC.17}
coincide so that the secret-key capacity of this channel is fully established.
This is also known as the PLOB bound~\cite{Pirandola.et.al.NC.17} and fully
characterizes the rate-loss scaling which affects any point-to-point QKD protocol.

One of the main tools used in Ref.~\cite{Pirandola.et.al.NC.17} was channel
simulation, where a quantum channel is simulated by applying an LOCC to a
suitable resource state. In particular, for the so-called teleportation
covariant channels, this simulation corresponds to
teleporting~\cite{Bennett.et.al.PRL.93} over the Choi matrix of the channel, a
property first noted for Pauli
channels~\cite{Bennett.et.al.PRA.96,Bowen.Bose.PRL.01}. Using this tool, one
can replace each transmission through a quantum channel with its simulation
and re-organize an adaptive (feedback-assisted) QKD protocol over the channel
into a much simpler block version. This technique is also known as
teleportation stretching and its combination with an entanglement measure as
the REE allows one to write simple single-letter upper bounds for the
secret-key capacity~\cite{Pirandola.et.al.NC.17}.

This methodology can be applied to bosonic Gaussian channels. In particular,
since these channels are teleportation-covariant, they can be simulated by
applying the CV teleportation
protocol~\cite{Vaidman.PRA.94,Braunstein.Kimble.PRL.98,Ralph.OL.99,Samtele2,telereview,Ralph.Lam.Polkinghorne.JOB.99}
over their asymptotic Choi matrices, as discussed in
Refs.~\cite{Pirandola.et.al.NC.17,Giedke.Cirac.PRA.02,Niset.Fiurasek.Cerf.PRL.09}%
. A bosonic Choi matrix is defined by propagating part of a two-mode squeezed
vacuum (TMSV) state~\cite{Weedbrook.et.al.RVP.12} through the channel, and
taking the limit of infinite energy. Therefore, the Choi matrix of a bosonic
channel is more precisely a limit over a succession of states. This also means
that a finite-energy simulation of a Gaussian channel, performed by
teleporting over a TMSV state, turns out to be imperfect with an associated
simulation error which must be carefully handled and propagated to the output
of adaptive protocols~\cite{Pirandola.et.al.NC.17,Pirandola.et.al.QST.18}.

An alternative way to simulate Gaussian channels is to implement the CV
teleportation protocol over a suitably-defined class of finite-energy Gaussian
states. This approach removes the limit of infinite energy in the resource
state, even though it remains at the level of the CV Bell detection, which is
defined as an asymptotic Gaussian measurement, whose limit realizes an ideal
projection onto displaced Einstein-Podolsky-Rosen (EPR) states. As shown in
Ref.~\cite{Scorpo.et.al.PRL.17,ScorpoERR,Scorpo.Adesso.SPIE.17}, it is
possible to realize such a finite-resource simulation. However, by combining
this type of channel simulation with the ingredients of
Ref.~\cite{Pirandola.et.al.NC.17}, i.e., teleportation stretching and REE, one
is not able to closely approximate the upper bounds to the secret key capacity
of bosonic Gaussian channels. This was shown in
Ref.~\cite{Laurenza.Braunstein.Pirandola.PRA.17} for the various
phase-insensitive Gaussian channels. Finite-resource simulation for the case
of the thermal-loss channel has also been considered in the numerical
investigation of Ref.~\cite{Kaur.Wilde.PRA.17}, where teleportation stretching
\cite{Pirandola.et.al.NC.17,Pirandola.et.al.QST.18} has been combined with
numerically-produced resource states to approximate the PLOB
thermal-loss\ upper bound~\cite{Pirandola.et.al.NC.17}.

More recently, in Ref.~\cite{Tserkis.Dias.Ralph.arxiv.18} all possible
resource states able to simulate a given Gaussian channel through
teleportation with finite resources were found analytically, and their
performance in terms of the entanglement of formation was studied. In this
work, we adopt this class of states, which can be parametrized in terms of
their {symplectic eigenvalues} and are optimized with respect to the REE.
Following the tools of Ref.~\cite{Pirandola.et.al.NC.17}, we therefore derive
corresponding upper bounds to the secret-key capacity of bosonic Gaussian
channels. Remarkably, these finite-energy upper bounds can be made as close as
possible to the infinite-energy bounds of Ref.~\cite{Pirandola.et.al.NC.17}
for all the phase-insensitive Gaussian channels, in particular, thermal-loss
channels, pure-loss channels, amplifiers, quantum-limited amplifiers, and
additive-noise Gaussian channels. Using the same class of states, we extend
the results from asymptotic security (infinite number of uses) to finite
number of uses, so that we can (approximately) bound the finite-size secret
key rates that are achievable by QKD\ protocols in the presence of loss and
thermal noise.

The paper is organized as follows. In Sec.~\ref{sec2}, we provide
preliminaries on Gaussian states, Gaussian channels, and the quantification of
entanglement via the REE. In Sec.~\ref{sec3}, we discuss the teleportation
simulation of Gaussian channels based on the new class of resource states. In
Sec.~\ref{sec4} we apply this tool to bound the secret-key capacity of the
phase-insensitive Gaussian channel, showing how our finite-energy bounds are
able to closely approximate the infinite-energy PLOB bounds. Sec.~\ref{sec5}
is for conclusions while Appendices~A and~B present tools and results for
finite-size bounds.

\section{Preliminaries\label{sec2}}

\subsection{Gaussian states}

Any n-mode bosonic state $\hat{\sigma}$ can be described by a vector of
quadrature field operators $\hat{q}:=(\hat{x}_{1},\hat{p}_{1},\ldots,\hat
{x}_{n},\hat{p}_{n})^{T}$, with $\hat{x}_{j}:=\hat{a}_{j}+\hat{a}_{j}^{\dag}$
and $\hat{p}_{j}:=i(\hat{a}_{j}^{\dag}-\hat{a}_{j})$, where $\hat{a}_{j}$ and
$\hat{a}_{j}^{\dag}$ are the annihilation and creation operators,
respectively, with commutator $[\hat{a}_{i}{,}{\hat{a}_{j}^{\dag}}]{=}%
\delta_{ij}$. Bosonic Gaussian states are those states which can be fully
characterized by the mean value and the variance of the quadratures $\hat{q}$.
In particular, a two-mode Gaussian state with zero mean value can be fully
described by a real and positive-definite matrix called the covariance matrix
(CM), {whose arbitrary element is defined by} $\sigma_{ij}=\frac{1}{2}%
\langle\{\hat{q}_{i},\hat{q}_{j}\}\rangle$, where $\{,\}$ is the
anticommutator~\cite{Serafini.B.17,Weedbrook.et.al.RVP.12,Adesso.Ragy.OSID.14}%
. In the standard or normal form, $\boldsymbol{\sigma}$ is given
by~\cite{Weedbrook.et.al.RVP.12,Duan.et.al.PRL.00,Simon.PRL.00}
\begin{equation}
\boldsymbol{\sigma}^{\text{sf}}=%
\begin{bmatrix}
a & 0 & c_{1} & 0\\
0 & a & 0 & c_{2}\\
c_{1} & 0 & b & 0\\
0 & c_{2} & 0 & b
\end{bmatrix}
\,. \label{sf}%
\end{equation}
Using symplectic transformations, $S$, any CM can be transformed into
$\boldsymbol{\nu}=S\boldsymbol{\sigma}S^{T}=\nu_{-}\mathds{1}\oplus\nu
_{+}\mathds{1}$, where $1\leq\nu_{-}\leq\nu_{+}$ are called symplectic
eigenvalues \cite{Serafini.Illuminati.DeSiena.JPB.04,Weedbrook.et.al.RVP.12}.
The purity of the state is given by $\mu=(\nu_{-}\nu_{+})^{-1}$.

\subsection{Gaussian channels}

Decoherence of quantum states is modeled through quantum channels which are
described by a completely positive trace-preserving map $\mathcal{C}$
\cite{Serafini.B.17,Weedbrook.et.al.RVP.12,Holevo.PIT.07}. Consider a two-mode
(zero-mean) Gaussian state with CM $\boldsymbol{\sigma}_{\text{in}}$. Assume
that the second mode is processed by a single-mode Gaussian channel
$\mathcal{G}$. Then, we have the following input-output transformation for the
CM%
\begin{equation}
\boldsymbol{\sigma}_{\text{in}}\overset{\mathcal{G}}{\longrightarrow
}\boldsymbol{\sigma}_{\text{out}}=(\mathds{1}\oplus\mathcal{U}%
)\boldsymbol{\sigma}_{\text{in}}(\mathds{1}\oplus\mathcal{U})^{T}%
+(0\oplus\mathcal{V}),
\end{equation}
where $\mathcal{U}=\sqrt{\tau}\mathds{1}$ represents the
attenuation/amplification operation and $\mathcal{V}=v\mathds{1}$ the induced
noise. Phase-insensitive Gaussian channels are the
following~\cite{Weedbrook.et.al.RVP.12,Holevo.PIT.07,Caruso,HolevoVittorio}:
(i) the thermal-loss channel $\mathcal{L}$ with transmissivity $0<\tau<1$ and
thermal noise $v=|1-\tau|(2\mean{n}+1)$, where $\mean{n}$ indicates the mean
number of photons of the environment (pure-loss channel or quantum-limited
attenuator for $\mean{n}=0$), (ii) the amplifier channel $\mathcal{A}$ with
gain $\tau>1$ and noise $v=|1-\tau|(2\mean{n}+1)$ (pure amplifier or
quantum-limited amplifier for $\mean{n}=0$), (iii) the additive-noise Gaussian
channel $\mathcal{N}$ with $\tau=1$ and added-noise variance $v>0$, and (iv)
the identity channel $\mathcal{I}$ with $\tau=1$ and $v=0$, representing the
ideal non-decohering channel. Note that we do not consider the conjugate of
the amplifier channel because it is entanglement-breaking and, therefore, has
zero secret-key capacity.

\subsection{Quantification of entanglement}

The \textit{bona fide} measure of entanglement for pure states is the entropy
of entanglement \cite{Bennett.et.al.PRA.96}, defined as $\mathcal{E}(\hat
{\rho}):=S(\mbox{tr}_{B}\hat{\rho})$, where $S(x):=-\mbox{tr}(x\log_{2}x)$ is
the von Neumann entropy, and $\mbox{tr}_{B}$ denotes the partial trace over
subsystem $B$ \cite{note}. For mixed states several measures have been defined
in the literature with different operational meanings
\cite{Plenio.Virmani.B.14,Horodecki.et.al.RMP.09,Adesso.Illuminati.JPAMT.07,Vidal.Werner.PRA.02}%
. In this work we use the REE~\cite{Vedral.et.al.PRL.97,Vedral.RMP.02} defined
by%
\begin{equation}
\mathcal{E}_{R}(\hat{\rho}):=\inf_{\hat{\rho}_{\text{sep}}}S(\hat{\rho}%
||\hat{\rho}_{\text{\textrm{sep}}})\,,
\end{equation}
where $\hat{\rho}_{\text{sep}}$ is an arbitrary separable state and
\begin{equation}
S(\hat{\rho}||\hat{\rho}_{\text{\textrm{sep}}}):=\mbox{tr}[\hat{\rho}(\log
_{2}\hat{\rho}-\log_{2}\hat{\rho}_{\text{\textrm{sep}}})]
\end{equation}
is the quantum relative entropy~\cite{Watrous,Hayashi,NielsenChuang}.

The REE has a geometrical interpretation as a \textquotedblleft
distance\textquotedblright\ between an entangled state and its closest
separable state. In general the computation of REE is a challenging task, and
thus we can calculate it only numerically. However, for Gaussian states an
upper bound of it can be defined by fixing a candidate separable state.
Specifically, for a Gaussian state $\hat{\rho}$ with CM $\boldsymbol{\rho}$ of
the form of Eq.~(\ref{sf}), we pick a separable state $\hat{\rho}_{\text{sep}%
}^{\ast}$ that has CM $\boldsymbol{\rho}_{\text{sep}}^{\ast}$, with the same
diagonal blocks as $\boldsymbol{\rho}$, but where the off-diagonal terms are
replaced as follows~~\cite[Supp. Note~4]{Pirandola.et.al.NC.17}%
\begin{equation}
c_{1,2}\rightarrow\pm\sqrt{(a-1)(b-1)}.
\end{equation}
Using the separable state $\hat{\rho}_{\text{sep}}^{\ast}$ we can then write
the upper bound
\begin{equation}
\mathcal{E}_{R}(\hat{\rho})\leq\mathcal{E}_{R}^{\ast}(\hat{\rho}):=S(\hat
{\rho}||\hat{\rho}_{\text{sep}}^{\ast})\,. \label{furtherUB}%
\end{equation}

The quantity $S(\hat{\rho}||\hat{\rho}_{\text{sep}}^{\ast})$ can be calculated
using the closed analytical formula derived in
Ref.~\cite{Pirandola.et.al.NC.17}, which is reviewed (and extended) in
Appendix~\ref{a:ree} and is based on the Gibbs representation for Gaussian
states~\cite{Banchi.Braunstein.Pirandola.PRL.15}. More precisely, for two
zero-mean Gaussian states with CMs $\boldsymbol{\rho}_{k}$ and
$\boldsymbol{\rho}_{\ell}$, their relative entropy is given by%
\begin{equation}
S(\hat{\rho}_{k}||\hat{\rho}_{\ell})=\Sigma(\boldsymbol{\rho}_{k}%
,\boldsymbol{\rho}_{\ell})-\Sigma(\boldsymbol{\rho}_{k},\boldsymbol{\rho}%
_{k})\,,
\end{equation}
where we have defined
\begin{equation}
\Sigma(\boldsymbol{\rho}_{k},\boldsymbol{\rho}_{\ell}):=\frac{\ln\det\left(
\frac{\boldsymbol{\rho}_{\ell}+i\boldsymbol{\Omega}}{2}\right)
+\mbox{tr}(\frac{\boldsymbol{\rho}_{k}\boldsymbol{G}_{\ell}}{2})}{2\ln2}\,,
\end{equation}
with $\boldsymbol{G}_{k}=2i\boldsymbol{\Omega}\coth^{-1}(i\boldsymbol{\rho
}_{k}\boldsymbol{\Omega})$~\cite{Banchi.Braunstein.Pirandola.PRL.15}, and the
matrix $\boldsymbol{\Omega}=\bigoplus_{i=1}^{2}\boldsymbol{\omega}$ is the
symplectic form, with $\boldsymbol{\omega}=%
\begin{bmatrix}
0 & 1\\
-1 & 0
\end{bmatrix}
$~\cite{Maths}.

\section{Finite-resource teleportation simulation\label{sec3}}

As discussed in Ref.~\cite{Pirandola.et.al.NC.17}, an arbitrary channel
$\mathcal{C}$ is called LOCC-simulable or $\hat{\rho}$-stretchable if it can
be simulated by a trace-preserving LOCC, $\Lambda$, and a suitable resource
state $\hat{\rho}$, i.e.%
\begin{equation}
\mathcal{C}(\hat{\sigma})=\Lambda(\hat{\sigma}\otimes\hat{\rho})\,.
\end{equation}
An important class is that of the Choi-stretchable channels, which can be
simulated via the Choi-state, defined as $\hat{\rho}^{\text{Choi}%
}:=\mathcal{I}\otimes\mathcal{C}(\hat{\varphi})$, with $\hat{\varphi}$ being
the maximally entangled state. This is always possible if $\mathcal{C}$ is
teleportation-covariant, i.e., it is covariant with respect to the random
unitaries of teleportation~\cite{Pirandola.et.al.NC.17}. In that case, the
resource state is its Choi matrix $\hat{\rho}^{\text{Choi}}$ and the LOCC
$\Lambda$ is teleportation.

As already mentioned before, bosonic Gaussian channels $\mathcal{G}$ are
teleportation-covariant, but their Choi matrices are asymptotic states. One
starts by considering a TMSV state $\hat{\varphi}_{\omega}$ with variance
$\omega=2\mean{n}+1$, with $\mean{n}$ being the mean number of photons in each
local mode. This is then partly propagated through $\mathcal{G}$ so as to
define its quasi-Choi matrix $\hat{\rho}_{\omega}^{\text{Choi}}:=\mathcal{I}%
\otimes\mathcal{G}(\hat{\varphi}_{\omega})$. Taking the limit for large
$\omega$, $\hat{\varphi}_{\omega}$ becomes the ideal EPR\ state, and
$\hat{\rho}_{\omega}^{\text{Choi}}$ defines the Choi matrix of $\mathcal{G}$.
Correspondingly, one may write the following asymptotic simulation for a
Gaussian channel
\begin{equation}
\mathcal{G}(\hat{\sigma})=\lim_{\omega}\Lambda(\hat{\sigma}\otimes\hat{\rho
}_{\omega}^{\text{Choi}})~, \label{asyCC}%
\end{equation}
where $\Lambda$ is the LOCC associated with CV teleportation~\cite{NoteBELL}.

\begin{figure}[t]
\centering \vspace{0.01cm}
\includegraphics[width=\columnwidth]{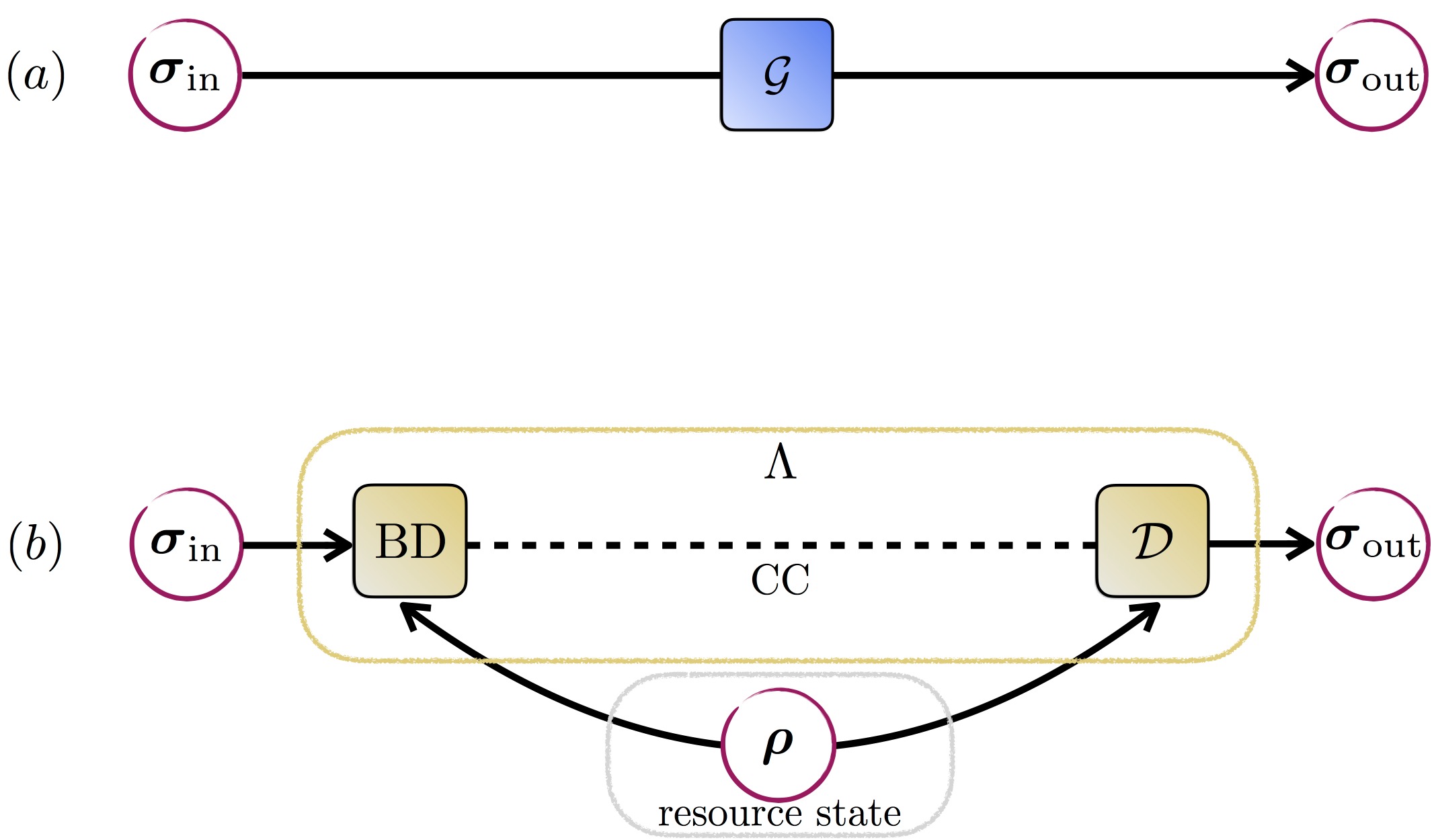}\caption{Finite-resource
simulation of bosonic Gaussian channels. In panel (a), we depict a
phase-insensitive Gaussian channel $\mathcal{G}=\mathcal{G}_{\tau,v}$
transforming the input state $\hat{\sigma}_{\text{in}}$ into the output state
$\hat{\sigma}_{\text{out}}$. In panel (b), we show its simulation by means of
a teleportation LOCC $\Lambda$. Its basic components are: (i) a CV\ Bell
detection (BD) between the input state $\hat{\sigma}_{\text{in}}$ and the
resource state $\hat{\rho}=\hat{\rho}_{\tau,v}$ as in
Eqs.~(\ref{physicalstates1})-(\ref{physicalstates}); (ii) the classical
communication (CC) of the Bell outcomes; and (iii) a conditional phase-space
displacement $D$ with suitable gain $\tau$~\cite{Braunstein.Kimble.PRL.98}
which provides the output teleported state $\hat{\sigma}_{\text{out}}$.}%
\label{fig1}%
\end{figure}

Generalizing previous ideas~\cite{Scorpo.et.al.PRL.17},
Ref.~\cite{Tserkis.Dias.Ralph.arxiv.18} has recently shown that an arbitrary
single-mode phase-insensitive Gaussian channel $\mathcal{G}=\mathcal{G}%
_{\tau,v}$, with parameters $\tau$ and $v$, can be simulated by CV
teleportation $\Lambda_{\tau}$ with gain $\tau$ over a suitable finite-energy
resource state $\hat{\rho}_{\tau,v}$ \cite{Tserkis.Dias.Ralph.arxiv.18}. In
other words, as also depicted in Fig.~\ref{fig1}, we may write
\begin{equation}
\mathcal{G}_{\tau,v}(\hat{\sigma})=\Lambda_{\tau}(\hat{\sigma}\otimes\hat
{\rho}_{\tau,v})~,
\end{equation}
where $\hat{\rho}_{\tau,v}$ is a zero-mean Gaussian state with CM
\begin{equation}
\boldsymbol{\rho}_{\tau,v}=%
\begin{bmatrix}
a & 0 & c & 0\\
0 & a & 0 & -c\\
c & 0 & b & 0\\
0 & -c & 0 & b
\end{bmatrix}
,
\end{equation}
where the elements of the CM\ are~\cite{Tserkis.Dias.Ralph.arxiv.18}%

\begin{align}
a  &  =\frac{|1-\tau|(\nu_{+}-\nu_{-})+(1+\tau)v-2\gamma}{(1-\tau)^{2}%
},\label{physicalstates1}\\
b  &  =\frac{\tau|1-\tau|(\nu_{+}-\nu_{-})+(1+\tau)v-2\gamma}{(1-\tau)^{2}%
},\label{physicalstates2}\\
c  &  =\frac{\tau|1-\tau|(\nu_{+}-\nu_{-})+2\tau v-(1+\tau)\gamma}{\sqrt{\tau
}(1-\tau)^{2}}, \label{physicalstates}%
\end{align}
and we have set~\cite{OtherSOL}
\begin{equation}
\gamma:=\sqrt{\tau(v-|1-\tau|\nu_{-})(v+|1-\tau|\nu_{+})}. \label{gamma}%
\end{equation}
Note that for $0<\tau<1$, we get states with $a\geq b$, while for $\tau>1$ we
get $a \leq b$. These elements are expressed in terms of the channel
parameters, $\tau$ and $v$, and may vary over the symplectic spectrum with the
constraints
\begin{equation}
1\leq\nu_{-}\leq2\mean{n}+1,\quad\quad\nu_{-}\leq\nu_{+}~, \label{range}%
\end{equation}
where $\mean{n}$ is the mean thermal number of the Gaussian channel
(thermal-loss or amplifier) \cite{noteresourcestates}.

For the special case of $\tau=1$, we have an additive-noise Gaussian channel
with added-noise variance $v>0$. In this case, taking the limit $\tau
\rightarrow1$ for the class in Eqs.~(\ref{physicalstates1}%
)-(\ref{physicalstates}) we get the following parametrization%

\begin{align}
a  &  =\frac{\nu_{-}^{2}+2\nu_{-}(\nu_{+}-v)+(\nu_{+}+v)^{2}}{4v}%
,\label{ADD0}\\
b  &  =\frac{\nu_{-}^{2}+2\nu_{-}(\nu_{+}+v)+(\nu_{+}-v)^{2}}{4v}%
,\label{ADD1}\\
c  &  =\frac{(\nu_{-}+\nu_{+}-v)(\nu_{-}+\nu_{+}+v)}{4v}, \label{ADD3}%
\end{align}
where $\nu_{-}\leq\nu_{+}$. In particular, by setting $\nu_{-}=\nu_{+}:=\nu$
in Eqs.~(\ref{ADD0})-(\ref{ADD3}) for the additive-noise Gaussian channel, we
may also consider a very simple single-parameter subclass of resource states
with%
\begin{equation}
a=b=\frac{\nu^{2}}{v}+\frac{v}{4},~c=\frac{\nu^{2}}{v}-\frac{v}{4}.
\label{ADDsimple}%
\end{equation}

\section{Secret-key capacity and bounds\label{sec4}}

\begin{figure}[t]
\centering
\includegraphics[width=\columnwidth]{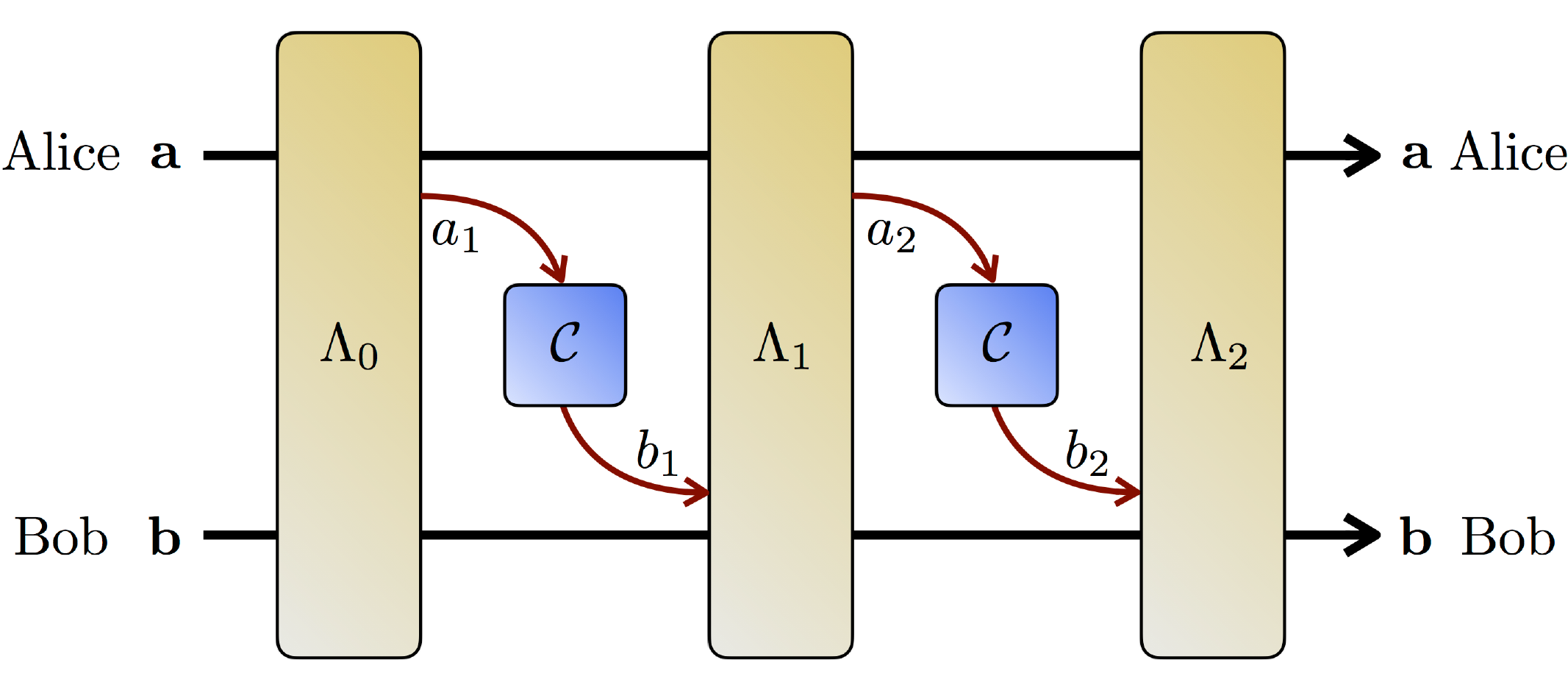} \caption{Schematic
description of an adaptive QKD protocol. In the first step, Alice and Bob
prepare the initial separable state $\hat{\rho}_{\mathbf{ab}}$ of their local
registers $\mathbf{a}$ and $\mathbf{b}$ by applying an adaptive LOCC
$\Lambda_{0}$. After the preparation of these registers, there is the first
transmission through the quantum channel $\mathcal{C}$. Alice picks a quantum
system from her local register $a_{1}\in\mathbf{a}$, which is therefore
depleted as $\mathbf{a}\rightarrow\mathbf{a}a_{1}$; then, system $a_{1}$ is
sent through the channel $\mathcal{C}$, with Bob getting the output $b_{1}$.
After transmission, Bob includes the output system $b_{1}$ in his local
register, which is augmented as $b_{1}\mathbf{b}\rightarrow\mathbf{b}$. This
is followed by Alice and Bob applying another adaptive LOCC $\Lambda_{1}$ to
their registers $\mathbf{a}$ and $\mathbf{b}$. In the second transmission,
Alice picks and sends another system $a_{2}\in\mathbf{a}$ through the quantum
channel $\mathcal{C}$ with output $b_{2}$ received by Bob. The remote parties
apply another adaptive LOCC $\Lambda_{2}$ to their registers and so on. This
procedure is repeated $n$ times, with the output state $\hat{\rho
}_{\mathbf{ab}}^{n}$ being finally generated for Alice's and Bob's local
registers.}%
\label{fig2}%
\end{figure}

The most general protocol for key generation is based on adaptive LOCCs (see
Fig.~\ref{fig2}). Each transmission through the quantum channel $\mathcal{C}$
is interleaved between two of such LOCCs. The general formalism goes as
follows. Assume that two remote users, Alice and Bob, have two local registers
of quantum systems (modes), $\mathbf{a}$ and $\mathbf{b}$, which are in some
fundamental state $\hat{\rho}_{\mathbf{a}}\otimes\hat{\rho}_{\mathbf{b}}$. The
two parties apply an adaptive LOCC $\Lambda_{0}$ before the first
transmission. In the first use of the channel, Alice picks a mode $a_{1}$ from
her register $\mathbf{a}$ and sends it through the channel $\mathcal{E}$. Bob
gets the output mode $b_{1}$ which is included in his local register
$\mathbf{b}$. The parties apply another adaptive LOCC $\Lambda_{1}$. Then,
there is the second transmission and so on. After $n$ uses, we have a sequence
of LOCCs $\{\Lambda_{0},\Lambda_{1},\ldots,\Lambda_{n}\}$ providing an output
state $\hat{\rho}_{\mathbf{ab}}^{n}$ which is $\epsilon$-close to a target
private state~\cite{Horodecki.et.al.PRL.05} with $nR_{n}^{\epsilon}$ bits.
This procedure characterizes an $(n,\epsilon,R_{n}^{\epsilon})$-protocol
$\mathcal{P}$. Taking the limit of large $n$, small $\epsilon$\ (weak
converse) and optimizing over $\mathcal{P}$, we define the secret-key capacity
of the channel $\mathcal{C}$ as%
\begin{equation}
K(\mathcal{C}):=\sup_{\mathcal{P}}\lim_{n,\epsilon}R_{n}^{\epsilon}\,.
\end{equation}

\begin{figure*}[tbh]
\centering
\includegraphics[scale=0.24]{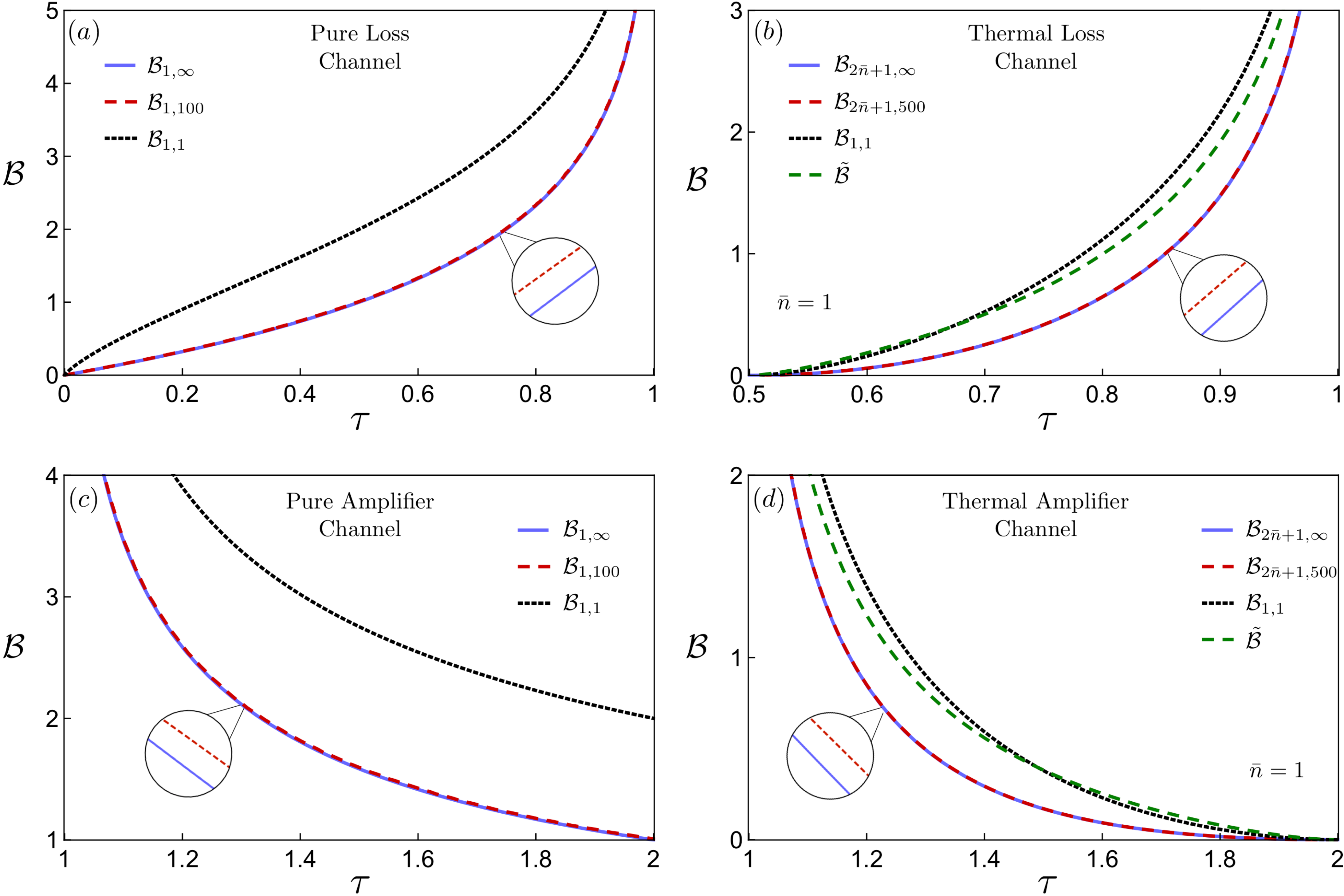}\caption{Upper bounds to the
secret-key rate capacity of lossy and amplifier channels (secret bits per
channel use versus transmissivity $0<\tau<1$ or gain $\tau>1$). In panels (a)
and (c) we show the results for pure loss and pure amplifier channels, while
panels (b) and (d) show the corresponding results for thermal loss and thermal
amplifier channels with $\mean{n}=1$. In the panels the lower blue line
indicates the infinite-energy bound {$\mathcal{B}_{2\mean{n}+1,\infty}$} of
Ref.~\cite{Pirandola.et.al.NC.17} while the green dashed line is the
approximate finite-energy bound $\tilde{\mathcal{B}}$ of
Ref.~\cite{Laurenza.Braunstein.Pirandola.PRA.17}, which is computed over the
class of states of Ref.~\cite{Scorpo.et.al.PRL.17}.{ The black dashed line
corresponds to our finite-energy bound $\mathcal{B}_{1,1}$ computed with a
pure resource state ($\nu_{\pm}=1$). Note that, for pure loss channels, this
bound $\mathcal{B}_{1,1}$ coincides with the previous finite-energy bound
given in \cite{Laurenza.Braunstein.Pirandola.PRA.17}. Then, the red dashed
line is our finite-energy bound $\mathcal{B}_{\nu_{-},\nu_{+}}$, computed with
$\nu_{-}=1$, $\nu_{+}=100$ for pure loss and pure amplifier channels, and with
$\nu_{-}=2\mean{n}+1$, $\nu_{+}=500$ for thermal loss and thermal amplifier
channels. As we see for increasing values of $\nu_{+}$, and thus increasing
simulation energy, we can approximate $\mathcal{B}_{2\mean{n}+1,\infty}$ as
closely as we want, while keeping the energy of the resource state finite
(although large).}}
\label{fig3}
\end{figure*}

Given a phase-insensitive Gaussian channel $\mathcal{G}=\mathcal{G}_{\tau,v}$
we may write its teleportation simulation by using our resource state
$\hat{\rho}=\hat{\rho}_{\tau,v}$ of Eqs.~(\ref{physicalstates1}%
)-(\ref{physicalstates}). Then, we may replace each transmission through the
channel by its simulation and stretch the adaptive protocol into a block
form~\cite{Laurenza.Braunstein.Pirandola.PRA.17,Pirandola.et.al.NC.17}, so
that we may write $\hat{\rho}_{\mathbf{ab}}^{n}=\Delta(\hat{\rho}^{\otimes
n})$ for a trace-preserving LOCC $\Delta$. Finally, we may upper bound the
secret-key capacity by computing the REE over the output state $\hat{\rho
}_{\mathbf{ab}}^{n}$. Since the REE is monotonic under $\Delta$ (data
processing) and sub-additive over tensor-products, we may
write~\cite{Laurenza.Braunstein.Pirandola.PRA.17,Pirandola.et.al.NC.17}%
\begin{equation}
K(\mathcal{G})\leq\mathcal{E}_{R}(\hat{\rho})\leq\mathcal{E}_{R}^{\ast}%
(\hat{\rho}), \label{plpl}%
\end{equation}
where $\mathcal{E}_{R}^{\ast}$ is defined according to Eq.~(\ref{furtherUB}).
{More precisely, the tightest bound is achieved by minimizing over the class
of resource states. Let us call $\mathcal{R}(\nu_{-},\nu_{+})$ the class of
states expressed by Eqs.~(\ref{physicalstates1})-(\ref{physicalstates}) [or by
Eqs.~(\ref{ADD0})-(\ref{ADD3}) in the limit }$\tau\rightarrow1${]. Then, for
any $\nu_{-}$ and $\nu_{+}$, we can consider the following bound
\begin{equation}
K(\mathcal{G})\leq\mathcal{B}_{\nu_{-},\nu_{+}}:=\min_{\hat{\rho}%
\in\mathcal{R}(\nu_{-},\nu_{+})}S(\hat{\rho}||\hat{\rho}_{\text{sep}}^{\ast}).
\label{finenbound}%
\end{equation}
}

We know that the minimum value of this bound is reached by the asymptotic Choi
matrix of the channel \cite{Pirandola.et.al.NC.17}. For thermal-loss and
thermal-amplifier channels this is retrieved for $\nu_{-}=2\mean{n}+1$ and
$\nu_{+}\rightarrow\infty$, while for additive-noise channels this corresponds
to $\nu_{\pm}\rightarrow\infty$. Thus, we can create monotonically decreasing
bounds in the following way: For thermal-loss and thermal-amplifier channels,
we set the lowest symplectic eigenvalue equal to $\nu_{-}=2\mean{n}+1$ and for
increasing $\nu_{+}$ (therefore simulation energy) we monotonically approach
the minimum value~\cite{Pirandola.et.al.NC.17} obtained for $\nu
_{+}\rightarrow\infty$, i.e.,
\begin{equation}
\mathcal{B}_{2\mean{n}+1,\nu_{+}}\gtrsim\mathcal{B}_{2\mean{n}+1,\infty}%
:=\lim_{\nu_{+}\rightarrow\infty}\mathcal{B}_{2\mean{n}+1,\nu_{+}}.
\label{limitB}%
\end{equation}
For additive-noise channels, we set $\nu_{-}=\nu_{+}:=\nu$ and for increasing
$\nu$ (therefore simulation energy) we monotonically approach the minimum
value~\cite{Pirandola.et.al.NC.17} for $\nu\rightarrow\infty$, i.e.,
\begin{equation}
\mathcal{B}_{\nu,\nu}\gtrsim\mathcal{B}_{\infty,\infty}:=\lim_{\nu
\rightarrow\infty}\mathcal{B}_{\nu,\nu}. \label{limitB2}%
\end{equation}
In the next section, we explicitly show the behaviour of these bounds for the
various Gaussian channels.

\section{Bounds for bosonic Gaussian channels}

\subsection{Thermal-loss channels}

Recall that a thermal-loss channel $\mathcal{L}$ can be modeled as a
beam-splitter operation $\exp[\theta(\hat{a}^{\dag}\hat{b}-\hat{a}\hat
{b}^{\dag})]$ with transmissivity $\tau=\cos^{2}\theta$, which mixes the input
state together with an environmental thermal state with variance
$2\mean{n}+1$. It is a pure-loss channel $\mathcal{L}_{p}$ for $\mean{n}=0$.
As shown in Ref.~\cite{Pirandola.et.al.NC.17}, the secret-key capacity of the
thermal-loss channel $\mathcal{L}$ is upper bounded by%
\begin{align}
K(\mathcal{L})  &  \leq{\mathcal{B}_{2\mean{n}+1,\infty}(\mathcal{L}%
)}\label{aaa1}\\
&  =\left\{
\begin{array}
[c]{c}%
-\log_{2}[(1-\tau)\tau^{\bar{n}}]-h(\bar{n})\quad\text{for}\quad\bar{n}%
<\frac{\tau}{1-\tau},\\
\quad\quad\quad\quad\quad0\hfill\text{otherwise,}%
\end{array}
\right. \nonumber
\end{align}
where we set $h(x):=(x+1)\log_{2}(x+1)-x\log_{2}x$. For the pure-loss channel
$\mathcal{L}_{p}$ we have the exact formula~\cite{Pirandola.et.al.NC.17}
\begin{equation}
K(\mathcal{L}_{p})={\mathcal{B}_{1,\infty}(\mathcal{L}_{p})}=-\log_{2}%
(1-\tau).
\end{equation}

Let us now compute the bound {$\mathcal{B}_{\nu_{-},\nu_{+}}$} for a
thermal-loss channel by fixing {$\nu_{-}=2\mean{n}+1$} and increasing
{$\nu_{+}$}. As shown in Fig.~\ref{fig3}, the finite-resource bound
{$\mathcal{B}_{2\mean{n}+1,\nu_{+}}$ rapidly approaches $\mathcal{B}%
_{2\mean{n}+1,\infty}(\mathcal{L})$ for increasing $\nu_{+}$} and this
approximation can be made as close as needed thanks to Eq.~(\ref{limitB}). In
Fig.~\ref{fig3}, we also show the corresponding performance for a pure-loss
channel $\mathcal{L}_{p}$. In Appendix~B, we provide further details on QKD
over a thermal-loss channel, showing how to bound the key rate $R_{n}%
^{\epsilon}$ of protocols with $\epsilon$-security and implemented a finite
number $n$ of times over the channel.\begin{figure}[t]
\centering
\includegraphics[width=\columnwidth]{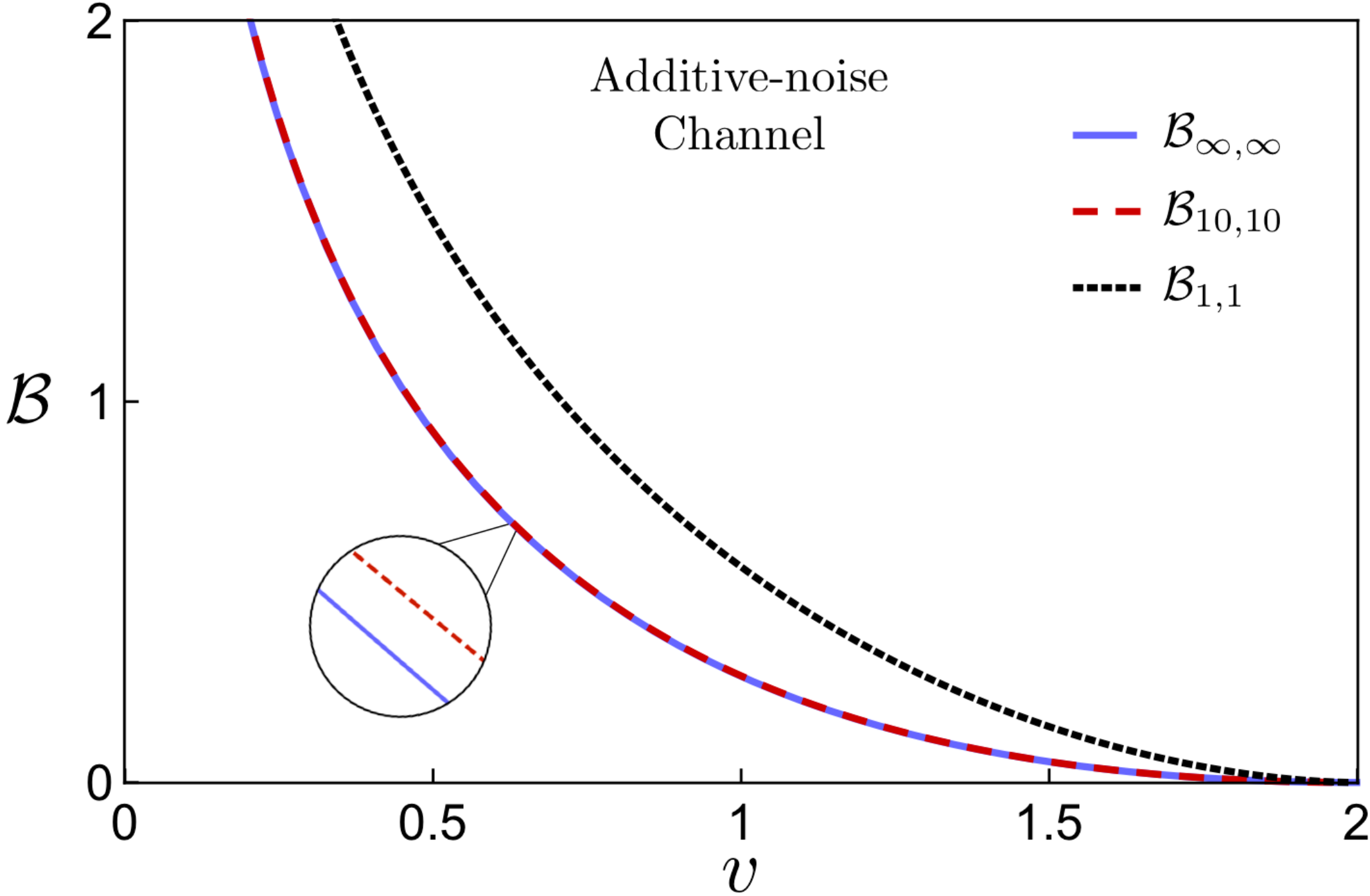} \caption{Upper bounds to
the secret-key capacity of the additive-noise Gaussian channel (secret bits
per channel use versus added noise $v$). The lower blue line indicates the
infinite-energy bound {$\mathcal{B}_{\infty,\infty}$ of
Ref.~\cite{Pirandola.et.al.NC.17}. Then, we show our improved finite-energy
bound $\mathcal{B}_{\nu,\nu}$ which is plotted for pure resource state, i.e.,
$\nu=1$ (black dashed line) and for $\nu=10$ (red dashed line). Note that the
previous bound given in \cite{Laurenza.Braunstein.Pirandola.PRA.17} coincides
with our finite-bound $\mathcal{B}_{1,1}$. For increasing values of $\nu$ we
can approximate $\mathcal{B}_{\infty,\infty}$} as closely as we want, while
keeping the energy of the resource state finite (despite being large).}%
\label{fig4}%
\end{figure}

\subsection{Quantum amplifiers}

A quantum amplifier channel $\mathcal{A}$ can be modeled by a two-mode
squeezing operation $\exp[r(\hat{a}\hat{b}-\hat{a}^{\dag}\hat{b}^{\dag})/2]$
with gain $\tau=\cosh r$, where $r$ is the squeezing
parameter~\cite{Caves.PRD.82}, which is applied to the input state together
with an environmental thermal state with $\bar{n}$ mean photons. In general,
for a thermal amplifier $\mathcal{A}$, we may write the following
infinite-energy bound~\cite{Pirandola.et.al.NC.17}%
\begin{align}
K(\mathcal{A})  &  \leq{\mathcal{B}_{2\mean{n}+1,\infty}(\mathcal{A})}\\
&  =\left\{
\begin{array}
[c]{c}%
-\log_{2}\left(  \frac{\tau-1}{\tau^{\bar{n}+1}}\right)  -h(\bar{n}%
)\quad\text{for}\quad\bar{n}<(\tau-1)^{-1},\\
\quad\quad\quad\quad\quad0\hfill\text{otherwise.}%
\end{array}
\right. \nonumber
\end{align}

For $\bar{n}=0$, we have a pure amplifier $\mathcal{A}_{p}$ and its secret-key
capacity is exactly known~\cite{Pirandola.et.al.NC.17}%
\begin{equation}
K(\mathcal{A}_{p})={\mathcal{B}_{1,\infty}(\mathcal{A}_{p})}=-\log_{2}%
(1-\tau^{-1}).
\end{equation}
By repeating the previous calculations, we may optimize over the class of
Eqs.~(\ref{physicalstates1})-(\ref{physicalstates}) at fixed {$\nu
_{-}=2\mean{n}+1$}. In Fig.~\ref{fig3} we see that {for increasing $\nu_{+}$,
we can approximate $\mathcal{B}_{2\mean{n}+1,\infty}(\mathcal{A})$ and
$\mathcal{B}_{1,\infty}(\mathcal{A}_{p})$ as much as we want.}

\subsection{Additive-noise Gaussian channel}

An additive-noise Gaussian channel $\mathcal{N}$ can be described as an
asymptotic case of either loss or thermal channels where $\tau\approx1$ and a
highly thermal state, i.e., classical, at the environmental input. It is known
that its secret-key capacity is upper-bounded as
follows~\cite{Pirandola.et.al.NC.17}%
\begin{align}
K(\mathcal{N})  &  \leq{\mathcal{B}_{\infty,\infty}(\mathcal{N})}\\
&  =\left\{
\begin{array}
[c]{c}%
\frac{v-2}{2\ln2}-\log_{2}(v/2)\quad\text{for}\quad v<2,\\
\quad\quad\quad\quad\quad0\hfill\text{otherwise.}%
\end{array}
\right. \nonumber
\end{align}

Here we assume the class specified by Eqs.~(\ref{ADDsimple}) for {increasing
values of $\nu$}. The corresponding finite-energy bound {$\mathcal{B}_{\nu
,\nu}(\mathcal{N})$ well-approximates the infinite-energy bound $\mathcal{B}%
_{\infty,\infty}(\mathcal{N})$}, as shown in Fig.~\ref{fig4}.

\section{Conclusions\label{sec5}}

In this work, we have improved the finite-energy upper bounds to the
secret-key capacities of one-mode phase-insensitive Gaussian channels.\ In
particular, we have shown that our finite-energy bounds can be made as close
as wanted to the infinite-energy bounds\ of Ref.~\cite{Pirandola.et.al.NC.17}.
This is possible because we are employing the general class of resource states
recently derived in Ref.~\cite{Tserkis.Dias.Ralph.arxiv.18}. This class
perfectly simulates Gaussian channels while it simultaneously allows us to
approach their asymptotic Choi matrices for {increasing energy}. For this
reason, we can always consider a perfect simulation with a finite-energy
resource state which can be made sufficiently close to the optimal one (i.e.,
the asymptotic Choi matrix).

Such an approach removes the need for using an asymptotic simulation at the
level of the resource state, even though the infinite energy limit still
remains at the level of Alice's quantum measurement which is ideally a
CV\ Bell detection (i.e., a projection onto displaced EPR states). Note that
our study regards point-to-point communication, but it can be immediately
extended to repeater chains and quantum networks~\cite{networkPIRS,Multipoint}%
. It would also be interesting to study the performance of the new class of
resource states in the setting of adaptive quantum metrology and quantum
channel discrimination~\cite{reviewMETRO,PirCo}, e.g., for applications in
quantum sensing~\cite{reviewSENSE}.

\section*{Acknowledgements}

This research has been supported by the Australian Research Council (ARC)
under the Centre of Excellence for Quantum Computation and Communication
Technology (CE170100012), the EPSRC via the \textquotedblleft UK Quantum
Communications Hub\textquotedblright\ (EP/M013472/1), and the European
Commission via `Continuous Variable Quantum Communications' (CiViQ, Project
ID: 820466).

\appendix

\section{Gaussian Relative Entropy and its Variance\label{a:ree}}

In this appendix we provide a self-contained proof of both the quantum
relative entropy between two arbitrary Gaussian
states~\cite{Pirandola.et.al.NC.17} and its variance~\cite{WildeREEVar}, that
were obtained using the techniques introduced
in~Refs.~\cite{Pirandola.et.al.NC.17,Banchi.Braunstein.Pirandola.PRL.15}.
Compared to the original derivations, the following proofs have the advantage
of being more compact and also more general, as they can be applied to
different notations available in the literature. Indeed, from bosonic creation
and annihilation operators we may define the bosonic quadrature operators
$\hat{x}_{j}=(\hat{a}_{j}+\hat{a}_{j}^{\dagger})/\sqrt{2\kappa}$ and $\hat
{p}_{j}=-i(\hat{a}_{j}-\hat{a}_{j}^{\dagger})/\sqrt{2\kappa}$ with different
normalizations $\kappa$, where the notation used in the main text is recovered
for $\kappa=1/2$, while the notation used
in~Refs.\cite{Pirandola.et.al.NC.17,WildeREEVar,Banchi.Braunstein.Pirandola.PRL.15}
is recovered with $\kappa=1$. The quadrature operators can be grouped into a
vector $\hat{q}:=(\hat{x}_{1},\hat{p}_{1},\dots,\hat{x}_{n},\hat{p}_{n})^{T}$
that satisfies the following commutation relations
\begin{equation}
\lbrack\hat{q},\hat{q}^{T}]=\frac{i\boldsymbol{\Omega}}{\kappa}.
\label{commrel}%
\end{equation}

Note that the operators $\kappa\hat{q}_{i}\hat{q}_{j}$ satisfy the same
algebraic properties of the operators defined for $\kappa=1$. As such we can
write any Gaussian state using the operator exponential
form~\cite{Banchi.Braunstein.Pirandola.PRL.15}
\begin{equation}
\hat{\rho}(\boldsymbol{\sigma},u)=\exp\left[  -\frac{\kappa}{2}(\hat{q}%
-u)^{T}\boldsymbol{G}(\hat{q}-u)\right]  /Z_{\rho}, \label{Gexpression}%
\end{equation}
where $u:=\langle\hat{q}\rangle_{\hat{\rho}}\in\mathbb{R}^{2n}$ is the first
moment,
\begin{equation}
Z_{\rho}=\det\left(  \kappa\boldsymbol{\sigma}+\frac{i\boldsymbol{\Omega}}%
{2}\right)  ^{1/2}, \label{PartitionFunction}%
\end{equation}
and the Gibbs matrix $\boldsymbol{G}$ is related to the CM $\boldsymbol{\sigma
}$ by
\begin{equation}
\boldsymbol{G}=2i\boldsymbol{\Omega}\,\coth^{-1}(2\kappa\boldsymbol{\sigma
}i\boldsymbol{\Omega}),~~\boldsymbol{\sigma}=\frac{1}{2\kappa}\coth\left(
\frac{i\boldsymbol{\Omega}\boldsymbol{G}}{2}\right)  i\boldsymbol{\Omega}.
\label{e.GtoV}%
\end{equation}
See also Ref.~\cite{Holev} for a similar treatment in different notation.

For calculating the relative entropy and its variance it is important to study
the expectation values of a generic quadratic operator $\hat q^{T}%
\boldsymbol{A}\hat q$, where $\boldsymbol{A}$ is a symmetric matrix. We focus
here on states with zero displacement $u=0$, as the generalization is
straightforward. The product of two operators can be expressed as
\begin{equation}
\hat q_{j} \hat q_{k} = \hat q_{j}\circ\hat q_{k} + [\hat q_{j},\hat q_{k}]/2
= \hat q_{j} \circ\hat q_{k} + \frac{i\Omega_{jk}}{2\kappa}~,
\end{equation}
where $\hat A\circ\hat B = (\hat A\hat B+\hat B\hat A)/2$, and we used the
commutation relations of Eq.~\eqref{commrel}. Since $\boldsymbol{\Omega}%
^{T}=-\boldsymbol{\Omega}$, for any symmetric $\boldsymbol{A}$ we may write
\begin{align}
\hat q^{T}\boldsymbol{A}\hat q  &  = \frac{i}{2\kappa}\mbox{tr}[\boldsymbol{A}%
\boldsymbol{\Omega}]+ \sum_{jk} \hat q_{j}\circ\hat q_{k} A_{jk}\\
&  = \sum_{jk} \hat q_{j}\circ\hat q_{k} A_{jk} ~. \label{quadsim}%
\end{align}
From the definition of the CM $\boldsymbol{\sigma}$ of a state $\hat\rho$ we
find then
\begin{equation}
\langle\hat q^{T}\boldsymbol{A}\hat q \rangle_{\hat\rho} =
\mbox{tr}[\boldsymbol{\sigma}\boldsymbol{A}]~. \label{e:mean}%
\end{equation}

For calculating the variance of the operator $\hat q^{T}\boldsymbol{A }\hat q$
we note that
\begin{align}
\label{quadexpand}\hat q^{T}\boldsymbol{A}\hat q\hat q^{T}\boldsymbol{A}\hat
q  &  = (\hat q^{T}\boldsymbol{A}\hat q) \circ(\hat q^{T}\boldsymbol{A}\hat q)
=\\
&  = \sum_{ijkl} A_{ij}A_{kl} (\hat q_{i}\circ\hat q_{j})\circ(\hat q_{k}%
\circ\hat q_{l})~,\nonumber
\end{align}
where Eq.~\eqref{quadsim} was used. In Ref.~\cite{MonrasWick} it has been
shown that
\begin{align}
\mbox{tr}[\hat\rho(\hat q_{i}\circ\hat q_{j})  &  \circ(\hat q_{k}\circ\hat
q_{l})] = \sigma_{ij}\sigma_{kl} + \sigma_{ik}\sigma_{jl} + \sigma_{il}%
\sigma_{jk} -\cr & - \frac1{4\kappa^{2}}\Omega_{ik}\Omega_{jl} -
\frac1{4\kappa^{2}}\Omega_{il}\Omega_{jk} ~.
\end{align}
Combining the above expression with Eq.~\eqref{quadexpand} we find
\begin{equation}
\langle(\hat q^{T}\boldsymbol{A}\hat q)^{2} \rangle_{\hat\rho} =
\mbox{tr}[\boldsymbol{\sigma}\boldsymbol{A}]^{2} + 2\mbox{tr}[\boldsymbol{A}%
\boldsymbol{\sigma}\boldsymbol{A}\boldsymbol{\sigma}] - \frac1{2\kappa^{2}%
}\mbox{tr}[\boldsymbol{A}\boldsymbol{\Omega}\boldsymbol{A}\boldsymbol{\Omega
}^{T}]~,
\end{equation}
and, in particular, the variance
\begin{align}
\langle(\hat q^{T}  &  \boldsymbol{A}\hat q- \langle\hat q^{T}\boldsymbol{A}%
\hat q\rangle_{\hat\rho})^{2} \rangle_{\hat\rho} = \cr & =
2\mbox{tr}[\boldsymbol{A}\boldsymbol{\sigma}\boldsymbol{A}\boldsymbol{\sigma}]
+ \frac1{2\kappa^{2}}\mbox{tr}[\boldsymbol{A}\boldsymbol{\Omega}%
\boldsymbol{A}\boldsymbol{\Omega}]~. \label{e:gvar}%
\end{align}

We are now ready to show how to compute the relative entropy $S(\hat{\rho}%
_{1}\Vert\hat{\rho}_{2})$ and its variance $V(\hat{\rho}_{1}\Vert\hat{\rho
}_{2})$, defined as
\begin{align}
S(\hat{\rho}_{1}\Vert\hat{\rho}_{2})  &  =\mbox{tr}[\hat{\rho}_{1}(\log
_{2}\hat{\rho}_{1}-\log_{2}\hat{\rho}_{2})]~,\label{RelEntr}\\
V(\hat{\rho}_{1}\Vert\hat{\rho}_{2})  &  =\mbox{tr}\left[  \hat{\rho}_{1}%
\hat{\Delta}(\hat{\rho}_{1},\hat{\rho}_{2})^{2}\right]  ~, \label{ReeVariance}%
\end{align}
where
\begin{equation}
\hat{\Delta}(\hat{\rho}_{1},\hat{\rho}_{2})=\log_{2}\hat{\rho}_{1}-\log
_{2}\hat{\rho}_{2}-S(\hat{\rho}_{1}\Vert\hat{\rho}_{2})~.
\end{equation}
Consider two generic Gaussian states $\hat{\rho}_{1}=\hat{\rho}%
(\boldsymbol{\sigma}_{1},u_{1})$ and $\hat{\rho}_{2}=\hat{\rho}%
(\boldsymbol{\sigma}_{2},u_{2})$. Without loss of generality we may define the
states $\tilde{\rho}_{1}=\hat{D}(u_{1})^{\dagger}\hat{\rho}(\boldsymbol{\sigma
}_{1},u_{1})\hat{D}(u_{1})=\hat{\rho}(\boldsymbol{\sigma}_{1},0)$ and
$\tilde{\rho}_{2}=\hat{D}(u_{1})^{\dagger}\hat{\rho}(\boldsymbol{\sigma}%
_{2},u_{2})\hat{D}(u_{1})=\hat{\rho}(\boldsymbol{\sigma}_{2},\delta)$, with
$\delta=u_{2}-u_{1}$, and where $\hat{D}(u)$ is the displacement
operator~\cite{Weedbrook.et.al.RVP.12}. Indeed, due to unitary invariance
$S(\tilde{\rho}_{1}\Vert\tilde{\rho}_{2})=S(\hat{\rho}_{1}\Vert\hat{\rho}%
_{2})$ and $V(\tilde{\rho}_{1}\Vert\tilde{\rho}_{2})=V(\hat{\rho}_{1}\Vert
\hat{\rho}_{2})$. From the exponential form of Eq.~\eqref{Gexpression} we
find
\begin{align}
-\log_{2}\tilde{\rho}_{1}  &  =\frac{2\ln Z_{\rho_{1}}+\kappa\hat{q}%
^{T}\boldsymbol{G}_{1}\hat{q}}{2\ln2}~,\\
-\log_{2}\tilde{\rho}_{2}  &  =\frac{2\ln Z_{\rho_{2}}+\kappa(\hat{q}%
-\delta)^{T}\boldsymbol{G}_{2}(\hat{q}-\delta)}{2\ln2}~,
\end{align}
so the relative entropy is obtained by taking the expectation value of the
above operators over $\tilde{\rho}_{1}\equiv\rho(\boldsymbol{\sigma}_{1},0)$.
Therefore, from Eqs.~\eqref{PartitionFunction} and~\eqref{e:mean}, and since
$\langle\hat{q}_{j}\rangle_{\tilde{\rho}_{1}}=0$, we may compute the entropic
functional
\begin{align}
\Sigma &  (\boldsymbol{\sigma}_{1},\boldsymbol{\sigma}_{j},\delta_j
):=-\mbox{tr}\left(  \tilde{\rho}_{1}\log_{2}\tilde{\rho}_{j}\right)
=\label{e:Sigma_functional}\\
&  =\frac{\ln\det\left(  \kappa\boldsymbol{\sigma}_{j}+\frac
{i\boldsymbol{\Omega}}{2}\right)  +\kappa\mbox{tr}(\boldsymbol{\sigma}%
_{1}\boldsymbol{G}_{j})+\kappa\delta_j^{T}\boldsymbol{G}_{j}\delta_j}{2\ln2},
\label{functional}%
\end{align}
where $\delta_1=0$ and $\delta_2=\delta$, 
from which we obtain the relative entropy~\eqref{RelEntr} as
\begin{equation}
S(\hat{\rho}_{1}\Vert\hat{\rho}_{2})=-\Sigma(\boldsymbol{\sigma}%
_{1},\boldsymbol{\sigma}_{1},0)+\Sigma(\boldsymbol{\sigma}_{1}%
,\boldsymbol{\sigma}_{2},\delta)~.
\label{RelENTROPY}
\end{equation}

The computation of the relative entropy variance is straightforward. In fact
we note that, from the exponential form in Eq.~\eqref{Gexpression} and the
relative entropy in Eqs.~\eqref{functional}-\eqref{RelENTROPY}, we may write
\begin{align}
\hat{\Delta}  &  =\log_{2}\tilde{\rho}_{1}-\log_{2}\tilde{\rho}_{2}%
-S(\tilde{\rho}_{1}\Vert\tilde{\rho}_{2})= &  & \nonumber\\
&  =\log_{2}Z_{2}-\log_{2}Z_{1}+\frac{\kappa(\hat{q}-\delta)^{T}%
\boldsymbol{G}_{2}(\hat{q}-\delta)}{2\ln2}-\frac{\kappa\hat{q}^{T}%
\boldsymbol{G}_{1}\hat{q}}{2\ln2}\cr & \phantom{=}~+\log_{2}Z_{1}-\log
_{2}Z_{2}-\frac{\kappa\mathrm{Tr}[\boldsymbol{\sigma}_{1}(\boldsymbol{G}%
_{2}-\boldsymbol{G}_{1})]}{2\ln2}-\frac{\kappa\delta^{T}\boldsymbol{G}%
_{2}\delta}{2\ln2}\cr  &  =\frac{\hat{q}^{T}(\boldsymbol{G}_{2}-\boldsymbol{G}%
_{1})\hat{q}-\mbox{tr}[\boldsymbol{\sigma}_{1}(\boldsymbol{G}_{2}%
-\boldsymbol{G}_{1})]-2\delta^{T}\boldsymbol{G}_{2}\hat{q}}{2\kappa^{-1}\ln
2}~,
\end{align}
where $\mbox{tr}[\boldsymbol{\sigma}_{1}(\boldsymbol{G}_{2}-\boldsymbol{G}%
_{1})]=\langle\hat{q}^{T}(\boldsymbol{G}_{2}-\boldsymbol{G}_{1})\hat{q}%
\rangle_{\tilde{\rho}_{1}}$. Since $\tilde{\rho}_{1}$ is a Gaussian state with
zero first moment, the expectation value of odd products of $\hat{q}$ is zero.
Therefore, the relative entropy variance is obtained from the variance of
$\hat{q}^{T}(\boldsymbol{G}_{2}-\boldsymbol{G}_{1})\hat{q}$, plus a correction
due to the displacement $\delta$. From Eqs.~\eqref{e:gvar} and~\eqref{e:mean}
the final result is then
\[
V(\hat{\rho}_{1}\Vert\hat{\rho}_{2})=\frac{4\kappa^{2}%
\mbox{tr}[\boldsymbol{\sigma}_{1}\tilde{G}\boldsymbol{\sigma}_{1}%
\boldsymbol{\tilde{G}}]+\mbox{tr}[\boldsymbol{\tilde{G}}\boldsymbol{\Omega
}\boldsymbol{\tilde{G}}\boldsymbol{\Omega}]+\delta^{T}\boldsymbol{B}\delta
}{2(2\ln2)^{2}}~,
\]
where $\boldsymbol{\tilde{G}}=\boldsymbol{G}_{1}-\boldsymbol{G}_{2}$,
$\delta=u_{1}-u_{2}$ and $\boldsymbol{B}=8\kappa^{2}\,\boldsymbol{G}%
_{2}\boldsymbol{\sigma}_{1}\boldsymbol{G}_{2}$.

\section{Finite-size bounds}

Besides bounding the (asymptotic) secret key capacity, we can use the
parametrization of resource states {$\mathcal{R}(\nu_{-},\nu_{+})$ [see
Eqs.~(\ref{physicalstates1})-(\ref{physicalstates}) and Eqs.~(\ref{ADD0}%
)-(\ref{ADD3})]} to bound the maximum finite-size key rate that is achievable
by an $(n,\epsilon,R_{n}^{\epsilon})$-protocol $\mathcal{P}$, i.e., a QKD
protocol which is implemented for a finite number $n$ of times with security
$\epsilon$. In fact, using channel simulation and teleportation stretching for
a bosonic Gaussian channel $\mathcal{G}$, one may easily
derive~\cite{Pirandola.et.al.NC.17,Pirandola.et.al.QST.18,WTB}%
\begin{equation}
K_{n,\epsilon}(\mathcal{G})\leq\frac{1}{n}D_{h}^{\epsilon}\left[  \hat{\rho
}^{\otimes n}||\left(  \hat{\rho}_{\text{sep}}^{\ast}\right)  ^{\otimes
n}\right]  , \label{bb1}%
\end{equation}
where $0<\epsilon<1$ and $D_{h}^{\epsilon}$ is the hypothesis testing relative
entropy~\cite{Li.AS.14}. Then, Ref.~\cite{Li.AS.14} directly provides
\begin{align}
D_{h}^{\epsilon}\left[  \hat{\rho}^{\otimes n}||\left(  \hat{\rho}%
_{\text{sep}}^{\ast}\right)  ^{\otimes n}\right]   &  =nS\left(  \hat{\rho
}||\hat{\rho}_{\text{sep}}^{\ast}\right) \label{bb2}\\
&  +\sqrt{nV\left(  \hat{\rho}||\hat{\rho}_{\text{sep}}^{\ast}\right)
}F(\epsilon)+O\left(  \log n\right)  ,\nonumber
\end{align}
where $F$ is the inverse of the cumulative Gaussian distribution, namely%
\begin{align}
F(\epsilon)  &  =\sup\{a\in\mathbb{R}~|f(a)\leq\epsilon\}~,\\
f(a)  &  =(2\pi)^{-1/2}\int_{-\infty}^{a}dx\exp(-x^{2}/2)~.
\end{align}

Combining Eqs.~(\ref{bb1}) and~(\ref{bb2}), it is immediate to write%
\begin{equation}
K_{n,\epsilon}(\mathcal{G})\leq S\left(  \hat{\rho}||\hat{\rho}_{\text{sep}%
}^{\ast}\right)  +\sqrt{\frac{V\left(  \hat{\rho}||\hat{\rho}_{\text{sep}%
}^{\ast}\right)  }{n}}F(\epsilon)+O\left(  \frac{\log n}{n}\right)  ,
\label{BBB}%
\end{equation}
as also used in Ref.~\cite{Kaur.Wilde.PRA.17}. The bound in Eq.~(\ref{BBB}) is
valid as long as the third moment is finite (e.g., see Ref.~\cite[Theorem~5]%
{Li.AS.14}), a condition that is certainly satisfied by energy-constrained
zero-mean Gaussian states. It is important to remark that the actual value of
the third moment has to be carefully considered in order to apply
Eq.~(\ref{BBB}) at small number of uses $n$. In other words, the scaling
$O\left(  n^{-1}\log n\right)  $ may actually be affected by a large
pre-factor, so that it becomes effective only for very large $n$. For this
reason, Eq.~(\ref{BBB}) has to be interpreted as an approximate bound when
applied to relatively small $n$.

\begin{figure}[t]
\centering \vspace{-0.3cm}
\includegraphics[width=\columnwidth]{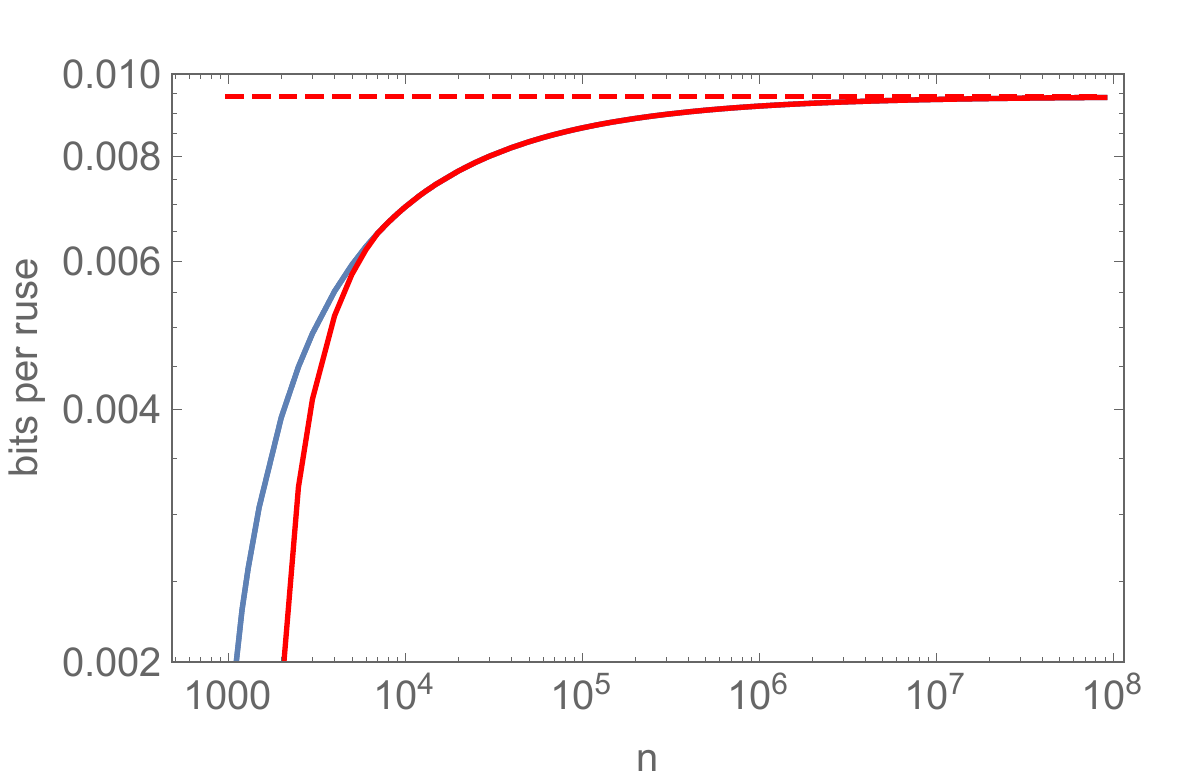}\caption{Secret-key
bits versus number $n$ of uses of a thermal-loss channel $\mathcal{L}$ with
transmissivity $\tau=0.01$ corresponding to $100$km of standard optical fiber
and thermal number$\ \bar{n}=0.0011$ corresponding to $\delta\simeq0.1$ excess
noise. We assume a security parameter $\epsilon=10^{-2}$. We plot the
optimized finite-size bound $\Phi_{n,\epsilon}(\mathcal{L})$ computed from
Eq.~(\ref{n2}) (solid red line) which approaches the asymptotic value
${\mathcal{B}_{2\mean{n}+1,\infty}(\mathcal{L})}$ of Eq.~(\ref{aaa1}) for
large $n$ (red dashed line). The optimal resource state $\hat{\rho}$ may have
low energy at finite $n$. For instance, at $n=2\times10^{3}$, this state has
spectrum $\nu_{-}\simeq1.001$ \ and $\nu_{+}\simeq3.33664$. For comparison, we
also plot the bound (solid blue line) that we would obtain with a resource
state of high energy, namely $\nu_{-}\simeq1.0022$ \ and $\nu_{+}%
\simeq3.99122\times10^{7}$.}%
\label{FIGfinite}%
\end{figure}

In general, we may consider a phase-insensitive Gaussian channel $\mathcal{G}$
and optimize the finite-size bound in Eq.~(\ref{BBB}) over the entire class of
resource states {$\mathcal{R}(\nu_{-},\nu_{+})$, which means to consider}%
\begin{align}
K_{n,\epsilon}(\mathcal{G})  &  \leq\Phi_{n,\epsilon}(\mathcal{G})+O\left(
\frac{\log n}{n}\right)  ,\label{n1}\\
\Phi_{n,\epsilon}(\mathcal{G})  &  :=\min_{\hat{\rho}\in\mathcal{R}(\nu
_{-},\nu_{+})}\left[  S\left(  \hat{\rho}||\hat{\rho}_{\text{sep}}^{\ast
}\right)  +\sqrt{\frac{V\left(  \hat{\rho}||\hat{\rho}_{\text{sep}}^{\ast
}\right)  }{n}}F(\epsilon)\right]  . \label{n2}%
\end{align}
As an example of application, we investigate a thermal-loss channel
$\mathcal{L}$ (similar results hold for the other phase-insensitive Gaussian
channels). Let us compute the finite-size optimized bound $\Phi_{n,\epsilon
}(\mathcal{L})$ for its $n$-use $\epsilon$-secure secret-key capacity
$K_{n,\epsilon}(\mathcal{L})$, assuming the numerical value $\epsilon=10^{-2}$
(smaller values can be considered but with further approximations, unless $n$
is of the order of $\epsilon^{-2}$). We then plot this approximate bound in
Fig.~\ref{FIGfinite}, showing its convergence for increasing $n$~\cite{Maths}.
In particular, the plot refers to a distance of $100$km in standard
optical-fiber at the loss rate of $0.2$dB/km and assumes an excess noise of
$\delta:=(1-\tau)\tau^{-1}\bar{n}\simeq0.1$. It is important to note that, at
finite $n$, the minimization in $\mathcal{R}(\nu_{-},\nu_{+})$ is taken for a
finite-energy resource state $\hat{\rho}$. The energy of this optimal resource
state then increases for increasing $n$.


\begin{thebibliography}{99}
\bibitem {note}Note that with the \textquotedblleft hat" symbol, e.g.
$\hat{\rho}$, we indicate density matrices while with bold $\boldsymbol{\rho}$
we indicate the corresponding covariance matrices.
\bibitem {NoteBELL}Note that, more precisely, one should write $\lim_{\omega
}\Lambda_{\omega}(\hat{\sigma}\otimes\hat{\rho}_{\omega}^{\text{Choi}})$,
where $\Lambda_{\omega}$ corresponds to a sequence of LOCCs, defined over a
finite-energy implementation of the ideal CV Bell
detection~\cite{Pirandola.et.al.NC.17,Pirandola.et.al.QST.18}.
\bibitem {OtherSOL}Note that another solution is given by replacing $\gamma$
with $-\gamma$ in Eqs.~(\ref{physicalstates1})-(\ref{physicalstates}). Also
note that, for a specific choice of the symplectic eigenvalues, we recover the
resource states of Ref.~\cite{Scorpo.et.al.PRL.17}.
\bibitem {noteresourcestates}Note that states with reversed symmetry for each
case, i.e., $a \leq b$ for $\tau<1$ and $a\geq b$ for $\tau>1$, can be
retrieved by interchanging $\nu_{-}$ and $\nu_{+}$, but then the corresponding
range is given by $1\leq\nu_{\pm}\leq2\mean{n}+1$.
\end{thebibliography}
\end{document}